\documentclass[%
 reprint,
showpacs,preprintnumbers,
amsmath,amssymb,
aps,
pra,
showkeys,
]{revtex4-1}

\usepackage{graphicx}
\usepackage{dcolumn}
\usepackage{bm}
\usepackage[pdftoolbar=true,pdfmenubar=true,colorlinks=true,linkcolor=red
,citecolor=blue,bookmarks=true,urlcolor=blue]{hyperref}

\usepackage{caption}
\usepackage{subcaption}

\usepackage{color}

\begin{document}
\preprint{
Preprint
}
\title{Effective three-band structure in Fe-based superconductors
}

\author{David M\"{o}ckli}
 \email{dmockli@if.uff.br}
\author{E.V.L. de Mello}
 \email{evandro@if.uff.br}

 \affiliation{Instituto de F\'{\i}sica da Universidade Federal Fluminense
 -- CEP 24.210-346 -- Niter\'{o}i -- RJ, Brazil}

\date{\today}

\begin{abstract}

We present self-consistent calculations of the multi-gap structure measured in 
some Fe-based superconductors. These materials are known to have structural
disorder in real space and a multi-gap structure
due to the $3d$ Fe-orbitals contributing to a complex Fermi surface topology
with hole and electron pockets. 
Different experiments identify three $s$-wave like superconducting gaps 
with a single critical temperature ($T_c$). We 
investigate the temperature dependence of these gaps by a 
multi-band Bogoliubov-deGennes theory at different
pockets in the presence of effective hybridizations between some bands
and an attractive temperature dependent intra-band interaction.
We show that this approach reproduces the three observed gaps and single $T_c$ in different compounds of Ba$_{1-x}$K$_{x}$Fe$_2$As$_2$, providing some insights on
the inter-band interactions.

\end{abstract}


\maketitle


\section{Introduction}

First principle
band theory calculations, such as local density approximation
(LDA) \cite{Singh2008,Ma2008,Xu2008a} on Ba$_{1-x}$K$_{x}$Fe$_2$As$_2$, 
predicted   five bands of the Fe 3d orbitals 
across the Fermi surface (FS) forming three hole-like
FSs centered at the zone center and two electron-like
FSs centered at the zone corner.  
This Fe-based superconductor (FeSC) revealed to be a highly complex high 
critical temperature ($T_c$) superconducting (SC) material with multi-band and multi-gap structure \cite{Ding2008}. Furthermore, it also possess an unusual
anisotropy in the ab-plane resistivity \cite{Chu2010,Tanatar2010,Ying2011}
just above $T_c$, possibly
related to an electronic phase separation transition \cite{Park2009} 
or a nematic phase \cite{Blomberg2013}.

Experiments such as angle-resolved photoemission spectroscopy (ARPES) \cite{Ding2008,Evtushinsky2009,Richard2011}, point-contact Andreev reflection spectroscopy (PCAR) \cite{Daghero2009,Szabo2009,Samuely2009a} and muon-spin rotation ($\mu$SR) \cite{Evtushinsky2009a} identify two or more nodeless $s$-wave like SC gaps. Here we want to address the classical ARPES experiment that
identified nearly two coinciding gap structures in the $\alpha$ and $\gamma$,
and another one in the large $\beta$ pocket of Ba$_{1-x}$K$_{x}$Fe$_2$As$_2$ with $T_c=38$~K \cite{Ding2008,Ding2011}. Differently than PCAR, ARPES is able to connect
the gaps with the pockets or bands in the Brillouin zone.

Earlier it was predicted by Suhl {\it et al} \cite{Suhl1959} that for the case in which two different bands develop SC gaps with different values, and in 
the absence of interband interaction, two $T_c$'s exist. A small amount of
interband scattering makes the two to merge to a single $T_c$. This result suggests
a clear interaction between the bands forming the hole and electron pockets  in
the Ba$_{1-x}$K$_{x}$Fe$_2$As$_2$ system. 
An additional difficulty to treat this system
is the multi-orbital character of the bands or pockets at the 
FS \cite{Hirschfeld2011} instead of the single structure like
the $s$-$d$ superconductors \cite{Suhl1959}. 

The main point tackled here concerns the $k$-dependent hybridization among the 
multi-bands forming the pockets where SC gaps were measured
by ARPES \cite{Ding2008,Ding2011}. Since the bands forming the pockets can be 
model by two \cite{Daghofer2010} or more orbitals \cite{Hirschfeld2011}, 
we describe the inter-band scattering by the hybridization of these 
orbitals. This generates a $k$-dependent interaction arising from a nonlocal 
character of the mixing \cite{Caixeiro2010}. In the real calculations we consider a special
case of a local constant multi-orbital hybridization between the $\alpha$-$\beta$
and $\beta$-$\gamma$ pockets, representing an average over the Brillouin zone.

\vspace{0.5cm}
\begin{figure}[h]
\centering
\captionsetup{justification=raggedright,
singlelinecheck=false
}
\begin{subfigure}[b]{0.23\textwidth}
\includegraphics[width=\textwidth]{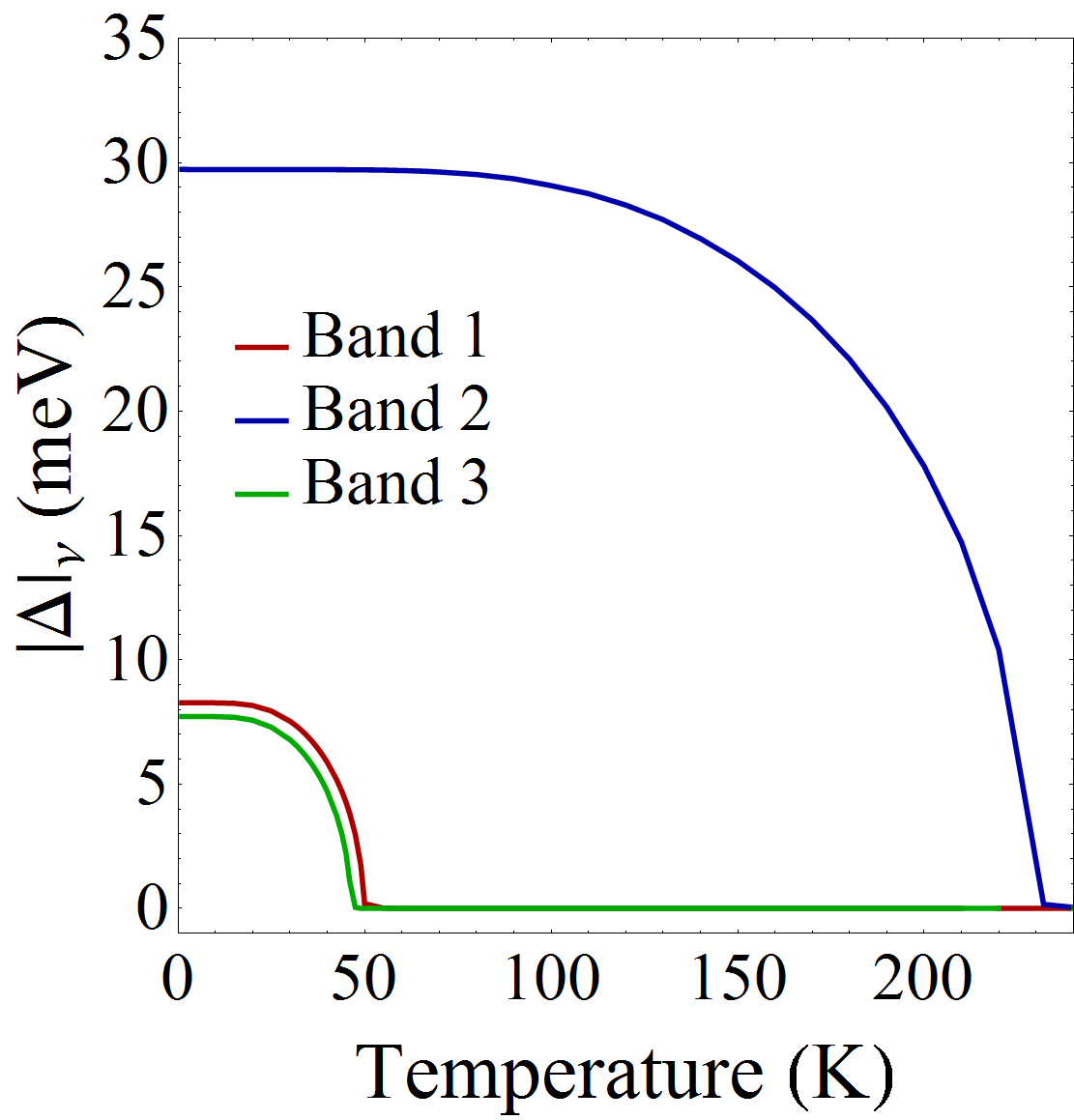}%
\end{subfigure}
\begin{subfigure}[b]{0.23\textwidth}
\includegraphics[width=\textwidth]{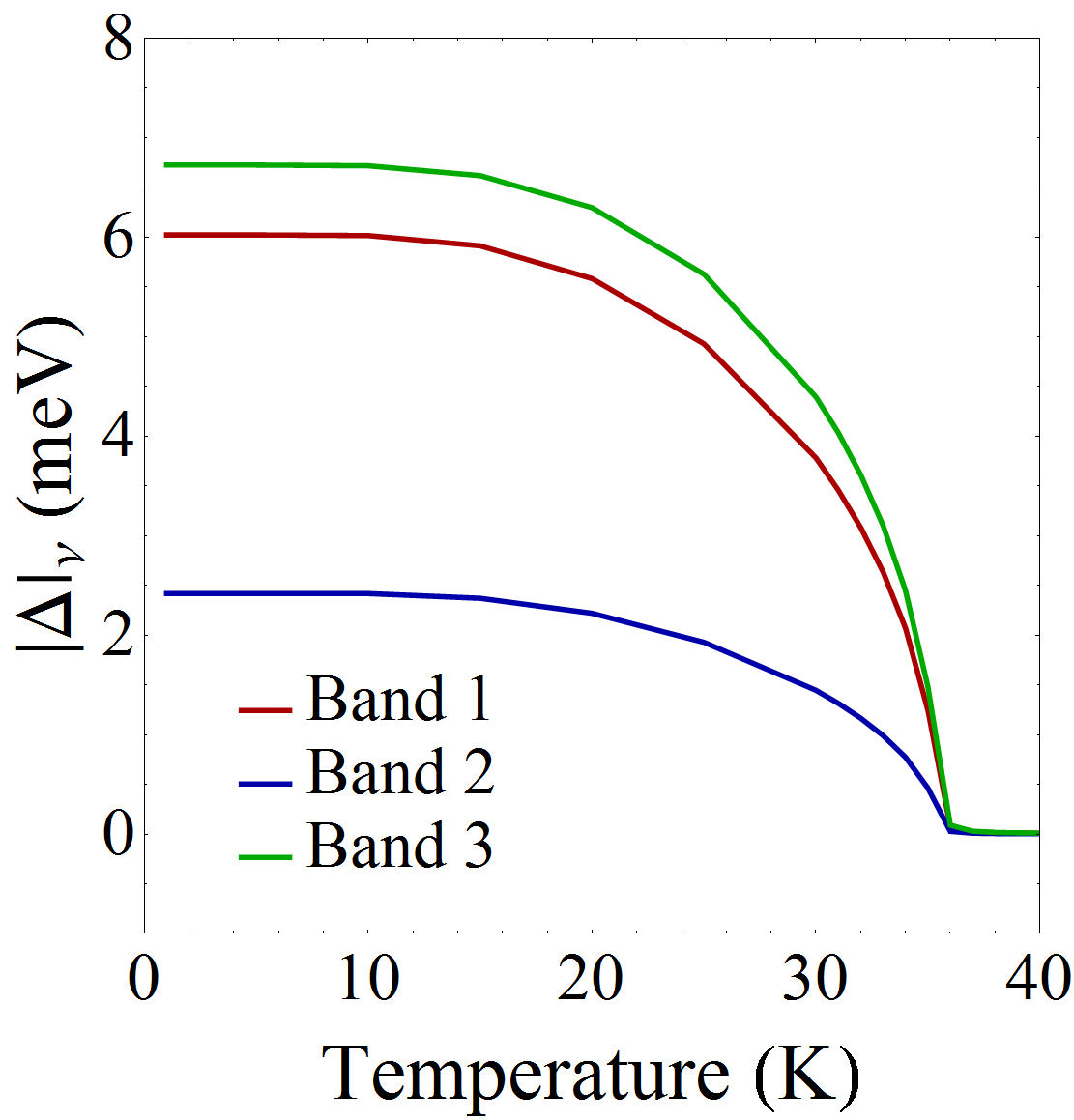}%
\end{subfigure}
\caption{
The left panel shows the SC gaps on three non-interacting bands with the 
same band parameters of the right panel extracted from \cite{Ding2008}
and \cite{Ding2011}. The big difference is the inter-band hopping.
Bands 1 (red), 2 (blue) and 3 (green) have the same hoppings as the $\alpha$, $\beta$ and $\gamma$ pockets respectively listed in table \ref{tab:tablehop}.
\label{hyb}
}%
\end{figure}

\section{Multi-band superconducting theory}

As mentioned, LDA calculations \cite{Singh2008,Ma2008,Xu2008a}
with the  Ba$_{1-x}$K$_{x}$Fe$_2$As$_2$ family
yielded five multi orbital bands across the FS. On the other hand, ARPES experiments \cite{Ding2008,Ding2011} measured three $s$-wave like SC
gaps at different pockets with quite different low temperature intensity and the
same $T_c$. This behavior is likely to occur due to interband 
interactions \cite{Suhl1959}. Thus, we propose the following mean field Hamiltonian,
with two terms, namely
\begin{equation} \label{mf}
\mathcal{H}_{\mathrm{MF}}=\mathcal{H}_{\mathrm{intra}}+\mathcal{H}_{\mathrm{inter}},
\end{equation}
where $\mathcal{H}_{\mathrm{intra}}$ contains the information of the three separate hole and electron pockets $\alpha$, $\beta$ and $\gamma$ \cite{Ding2008,Ding2011}, 

\begin{widetext}
\begin{equation} \label{intra}
\mathcal{H}_{\mathrm{intra}}=\sum_{\mathbf{i}\neq\mathbf{j}}\sum_{\nu=\alpha,\beta,\gamma} \sum_\sigma \, t_{\nu}(\mathbf{i},\mathbf{j}) c_{\nu\sigma}^\dag(\mathbf{i})c_{\nu\sigma}(\mathbf{j})
-\sum_{\mathbf{i},\nu,\sigma}\tilde{\mu}_\nu(\mathbf{i})c_{\nu\sigma}^\dag(\mathbf{i})c_{\nu\sigma}(\mathbf{i})
+\sum_{\mathbf{i}\mathbf{j},\nu}\left[\Delta_\nu(\mathbf{i},\mathbf{j})c_{\nu\uparrow}^\dag(\mathbf{j})c_{\nu\downarrow}^\dag (\mathbf{i})+\mathrm{h.c.} \right ].
\end{equation}
\end{widetext}
The $t_{\nu}(\mathbf{i},\mathbf{j})$ are intra-band hoppings between lattice sites $\mathbf{i}$ and $\mathbf{j}$  up to second nearest neighbors 
as derived by the ARPES band dispersion~\cite{Ding2011}, based on their
results and on the theory of magnetic excitation on a four orbital model~\cite{Korshunov2008}.
We introduced a short-hand notation for the shifted chemical potential $\tilde{\mu}_\nu(\mathbf{i})=\mu_\nu(\mathbf{i})-U_\nu(\mathbf{i})\rho_\nu(\mathbf{i})/2$,
which includes:
the local chemical potential $\mu_\nu(\mathbf{i})$, the on-site Coulomb repulsion $U_\nu(\mathbf{i})$,
that due to the mean-field treatment just enters as a rescaling factor,
and the band charge density $\rho_\nu(\mathbf{i})$. 
The spin index $\sigma$ assumes either $\uparrow$ or $\downarrow$, and the creation and annihilation operators obey the Fermi anti-commutation relation $\{ c_{\mu\sigma}(\mathbf{i}),c^\dag_{\nu\sigma^\prime}(\mathbf{j})\}=\delta_{\mathbf{i}\mathbf{j}}\delta_{\mu\nu}\delta_{\sigma\sigma^\prime}$.

The second term in equation \eqref{mf} includes the inter-band contribution along the same lines of the theory developed 
by Kishore {\it et al.} \cite{Kishore1970,Japiassu1992}, and Caixeiro {\it et al.}  \cite{Caixeiro2010}, where the hybridization 
between overlapping bands near the Fermi surface is approximated by
a constant (representing an average over the Brillouin zone) nearest
neighbour hopping
\begin{equation} \label{inter}
\mathcal{H}_{\mathrm{inter}}=\sum_{\mathbf{i}\neq\mathbf{j}}\sum_{\mu\neq\nu}\sum_\sigma t_{\mu\nu}(\mathbf{i},\mathbf{j}) \, c_{\mu\sigma}^\dag(\mathbf{i})c_{\nu\sigma}(\mathbf{j}).
\end{equation}

The band density $\rho_\nu(\mathbf{i})$ is self-consistently regulated by adjusting $\mu_\nu(\mathbf{i})$ until $\rho_\nu(\mathbf{i})$ converges to some fixed value.
Due to its local expression this approach can be applied to cases where there is 
some degree of disorder in the electronic density.

The local SC band gap is $\Delta_\nu(\mathbf{i},\mathbf{j})=-V(T)\langle c_{\nu\downarrow} (\mathbf{i}) c_{\nu\uparrow}(\mathbf{j})\rangle$,
where $\langle \cdots \rangle$ represents a thermal average. 
$V(T)$ is a temperature dependent potential derived from a two-phase separated
system. Using a typical Ginzburg-Landau free energy expansion it is
possible to show that free energy barrier between the two phases
is proportional to $(T_{\mathrm{PS}}-T)^2$, 
where $T_{\mathrm{PS}}$ is the phase separation critical temperature. 
This approach has been applied to cuprates and in this case 
$T_{\mathrm{PS}}$ is associated with the pseudogap temperature 
$T^*$ \cite{DeMello2009,Mello2012}. As mentioned, there is experimental evidence that 
the Ba$_{1-x}$K$_{x}$Fe$_2$As$_2$ system  may also undergo an electronic
phase separation \cite{Park2009} or nematic order \cite{Blomberg2013}.
Based on these observations, we take
\begin{equation}
V(T)=-|V_0|\left(1-\frac{T}{T_{\mathrm{PS}}} \right )^2.
\label{varying_temp}
\end{equation}

The Hamiltonian defined by \eqref{mf} is diagonalized by the unitary Bogoliubov transformation
\begin{equation} \label{2bogol}
c_{\nu\sigma}(\mathbf{i})=\sum_n\left(u_{ n\nu}(\mathbf{i})\gamma_{ n\sigma}-\mathrm{sgn}(\sigma)v_{ n\nu}^*(\mathbf{i})\gamma^\dag_{ n,-\sigma} \right ),
\end{equation}
which 
leads to the multi-band Bogoliubov-deGennes (BdG) equation
\begin{widetext}
\begin{equation} \label{multi_bdg}
E_{n}
\begin{bmatrix}
u_{ n\mu}(\mathbf{i})\\ 
v_{ n\mu}(\mathbf{i})
\end{bmatrix}=
\sum_{\mathbf{j}}
\begin{bmatrix}
\sum_\nu\left(  t_{\mu\nu}(\mathbf{i},\mathbf{j})- \tilde{\mu}_\nu(\mathbf{i})\delta_{\mu\nu}\right ) &\Delta_\mu(\mathbf{i},\mathbf{j})\delta_{\mu\nu}\\ 
\Delta^*_\mu(\mathbf{i},\mathbf{j})\delta_{\mu\nu} & -\sum_\nu\left(  t^*_{\mu\nu}(\mathbf{i},\mathbf{j})+\tilde{\mu}_\nu(\mathbf{i})\delta_{\mu\nu}\right )
\end{bmatrix}
\begin{bmatrix}
u_{ n\nu}(\mathbf{j})\\ 
v_{ n\nu}(\mathbf{j})
\end{bmatrix},
\end{equation}
\end{widetext}
where $t_{\nu\nu}\equiv t_{\nu}$ in equation \eqref{multi_bdg}. The full BdG matrix has dimension $6N^2\times 6N^2$ in the three-band case, where $N\times
 N$ is the lattice's dimension. The positive quasi-particle excitations $\{E_n\}$ are used to self-consistently calculate the temperature dependent
local $s$-wave gap for each band
\begin{equation}
\Delta_\nu(\mathbf{i})=-V_{\nu}(\mathbf{i})\sum_nu_{n\nu}(\mathbf{i})v_{n\nu}^*(\mathbf{i})
\tanh\left(\frac{E_n}{2k_BT}\right),
\label{swavegap}
\end{equation}
and the local band density $\rho_\nu(\mathbf{i})=\sum_{\sigma}\left\langle c^\dag_{\nu\sigma}(\mathbf{i})c_{\nu\sigma}(\mathbf{i})\right\rangle $ as
\begin{equation}
\rho_\nu(\mathbf{i})=2\sum_n\left(|u_{n\nu}(\mathbf{i})|^2 f_n+|v_{n\nu}(\mathbf{i})|^2 (1-f_n)\right ),
\label{density}
\end{equation}
where $f_n$ is the Fermi distribution of quasi-particles. 
Equation \eqref{swavegap} allows for the simultaneous determination of $\Delta_\alpha(T)$, $\Delta_\beta(T)$, and $\Delta_\gamma(T)$.

A typical solution of equation \eqref{swavegap} with a single constant attractive potential for all three bands with independent band dynamics (no inter-band hoppings) is shown on the left panel of figure \ref{hyb}. 
The intra-band hoppings are those derived from Ding {\it et al} \cite{Ding2011} 
that reproduces the three $\alpha$, $\beta$ and $\gamma$ pockets and are listed 
in table~\ref{tab:tablehop}. 

Notice that lower intra-band hoppings or lower kinetic energies yield bigger gaps and 
larger $T_c$'s. However, as predicted by Suhl {\it et al.} \cite{Suhl1959} for 
two bands, non-zero inter-band hoppings modify the three-gap structure drastically as shown on the right panel of figure \ref{hyb}. The crucial difference in the second case is a single value for $T_c$ for all three bands, in accordance with experimental observations
\cite{Ding2008,Evtushinsky2009,Richard2011,Daghero2009,Szabo2009,Samuely2009a}.

\section{Temperature dependence of the three-gap structure}
\begin{figure}[h]
\centering
\captionsetup{justification=raggedright,
singlelinecheck=false
}
\begin{subfigure}[b]{0.23\textwidth}
\includegraphics[width=\textwidth]{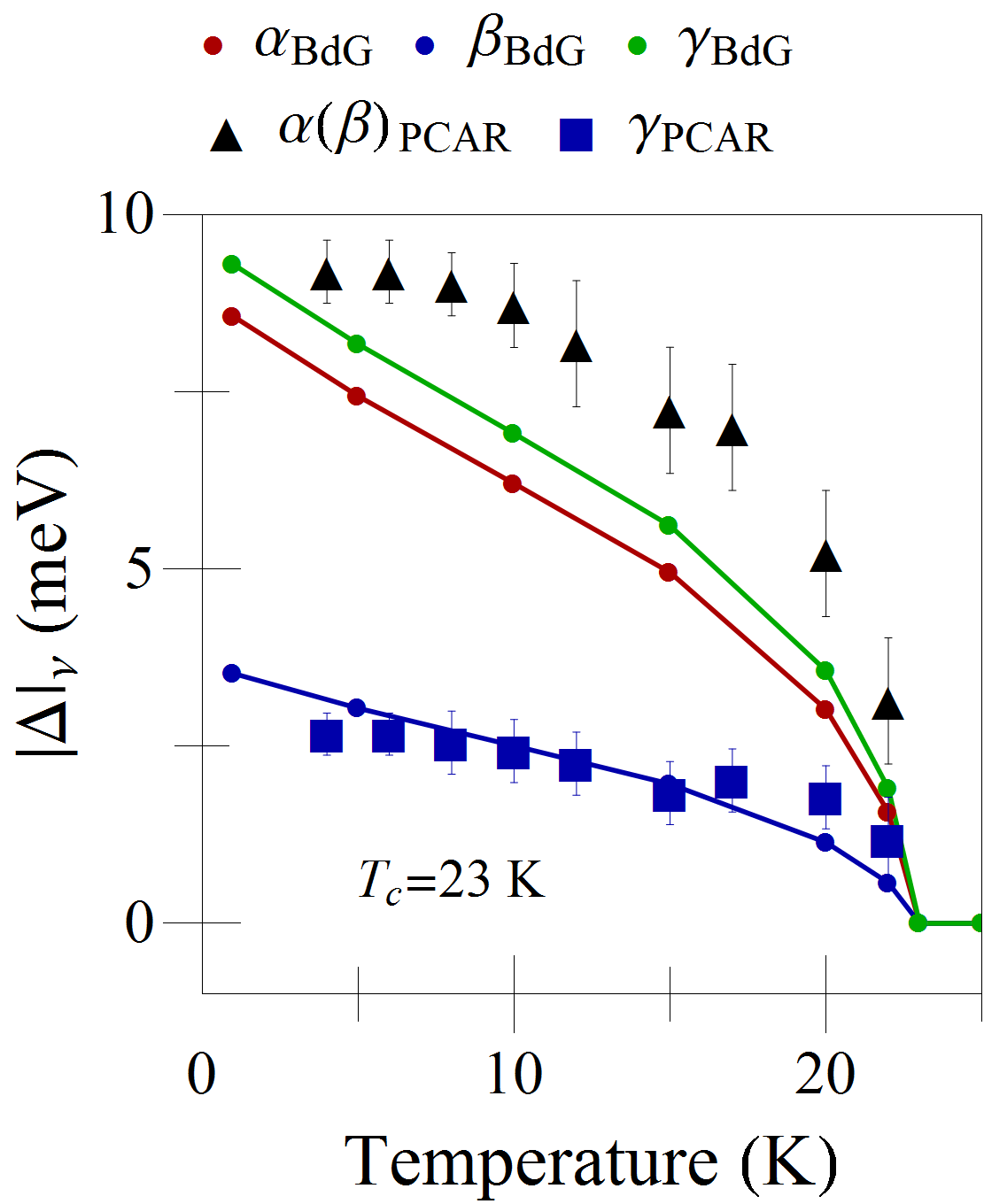}%
\end{subfigure}
\begin{subfigure}[b]{0.23\textwidth}
\includegraphics[width=\textwidth]{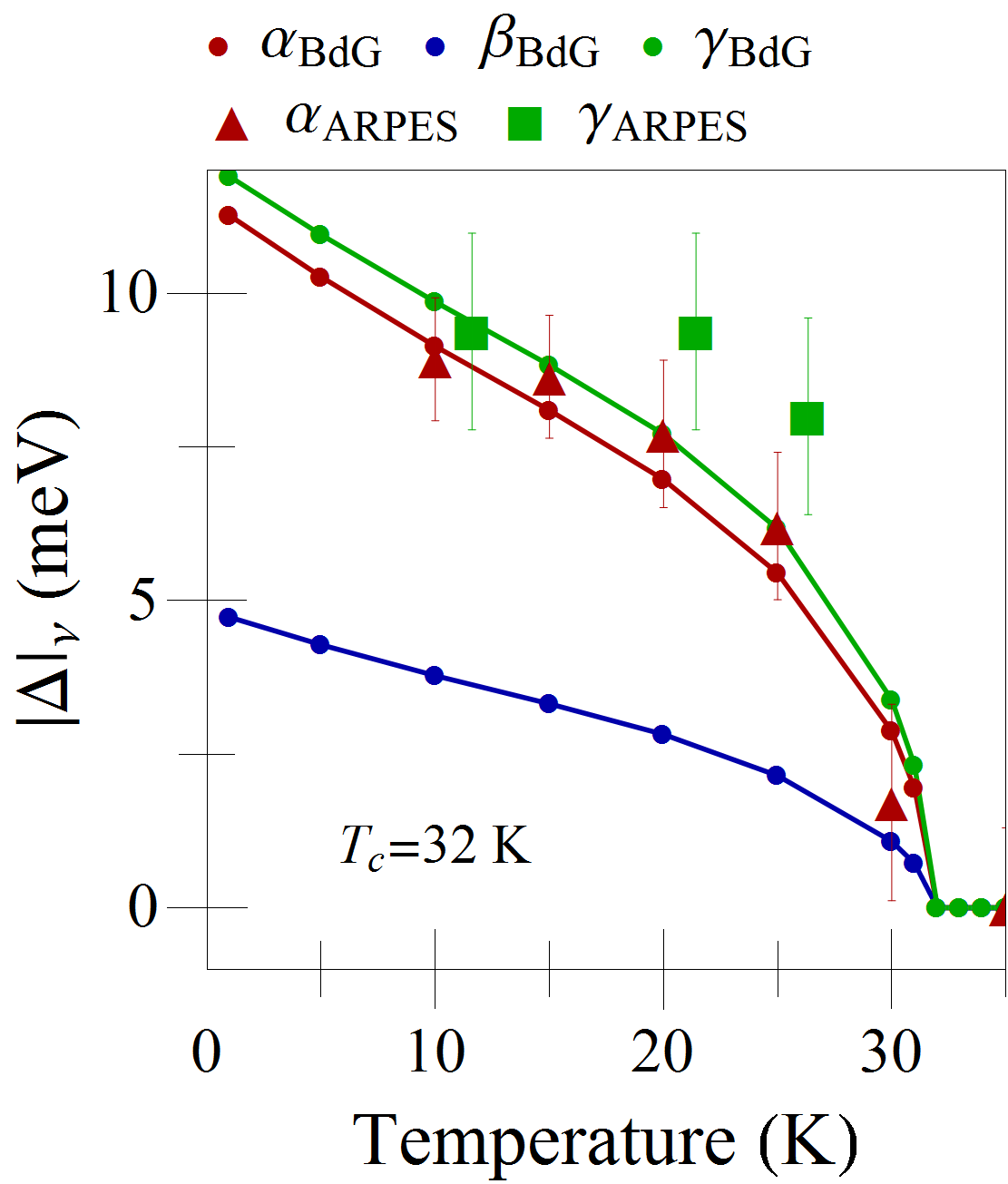}%
\end{subfigure}
\caption{
 On the left panel we show the theoretical BdG curves compared to PCAR \cite{Szabo2009,Samuely2009a} for Ba$_{1-x}$K$_{x}$Fe$_2$As$_2$ with $T_c=23$ K. PCAR identifies two SC bands that may correspond to the nearly coinciding $\alpha$ 
 and $\gamma$ pocket identified by ARPES. On the right panel we show the BdG curves compared to the three bands from ARPES \cite{Evtushinsky2009} for the compound with $T_c=32$ K. Reference \cite{Evtushinsky2009} only estimates gap values for the $\beta$ pocket. \label{tc2332}
}%
\end{figure}

\begin{figure*}
\centering
\captionsetup{justification=raggedright,
singlelinecheck=false
}
\begin{subfigure}[b]{0.24\textwidth}
\includegraphics[width=\textwidth]{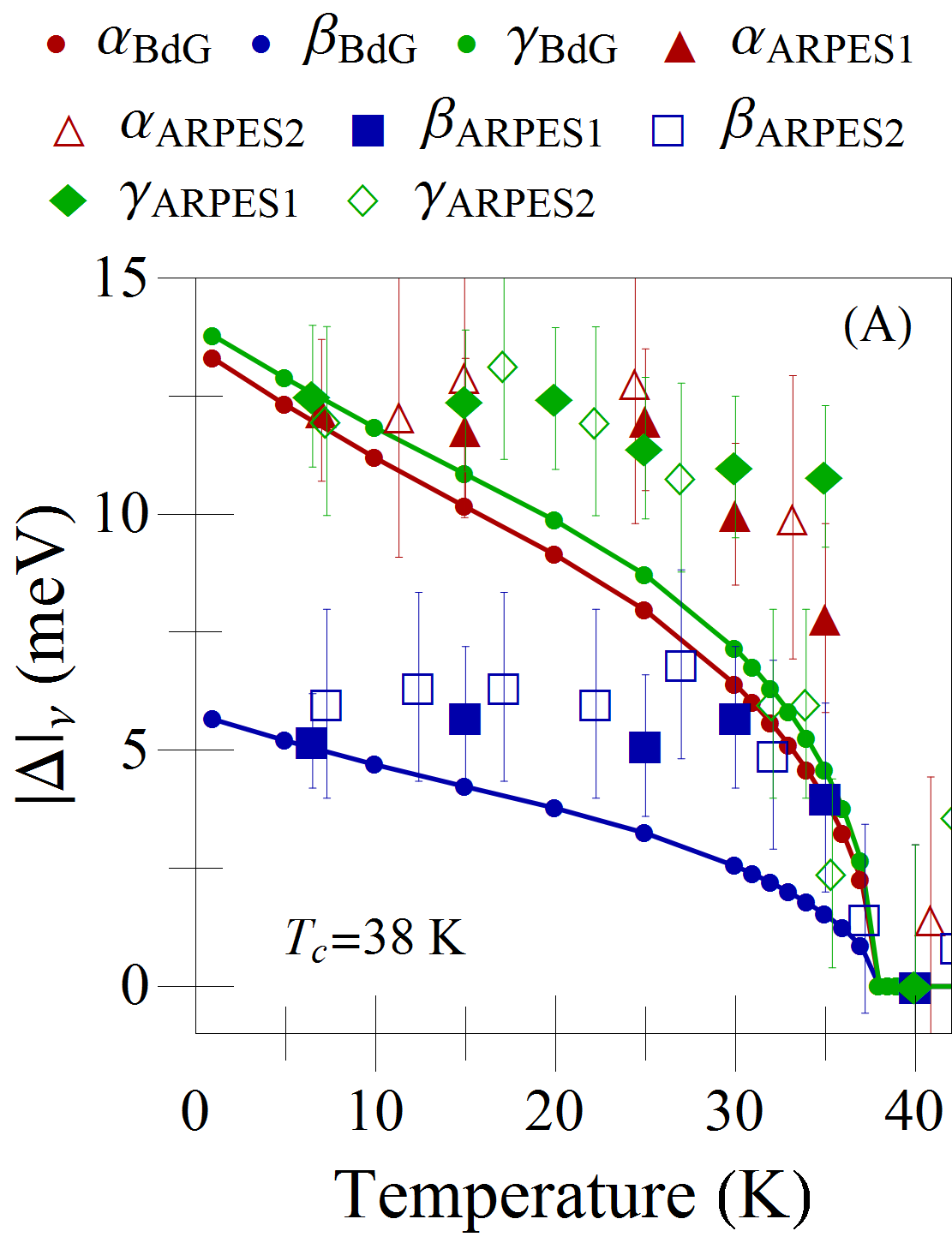}%
\end{subfigure}
\begin{subfigure}[b]{0.24\textwidth}
\includegraphics[width=\textwidth]{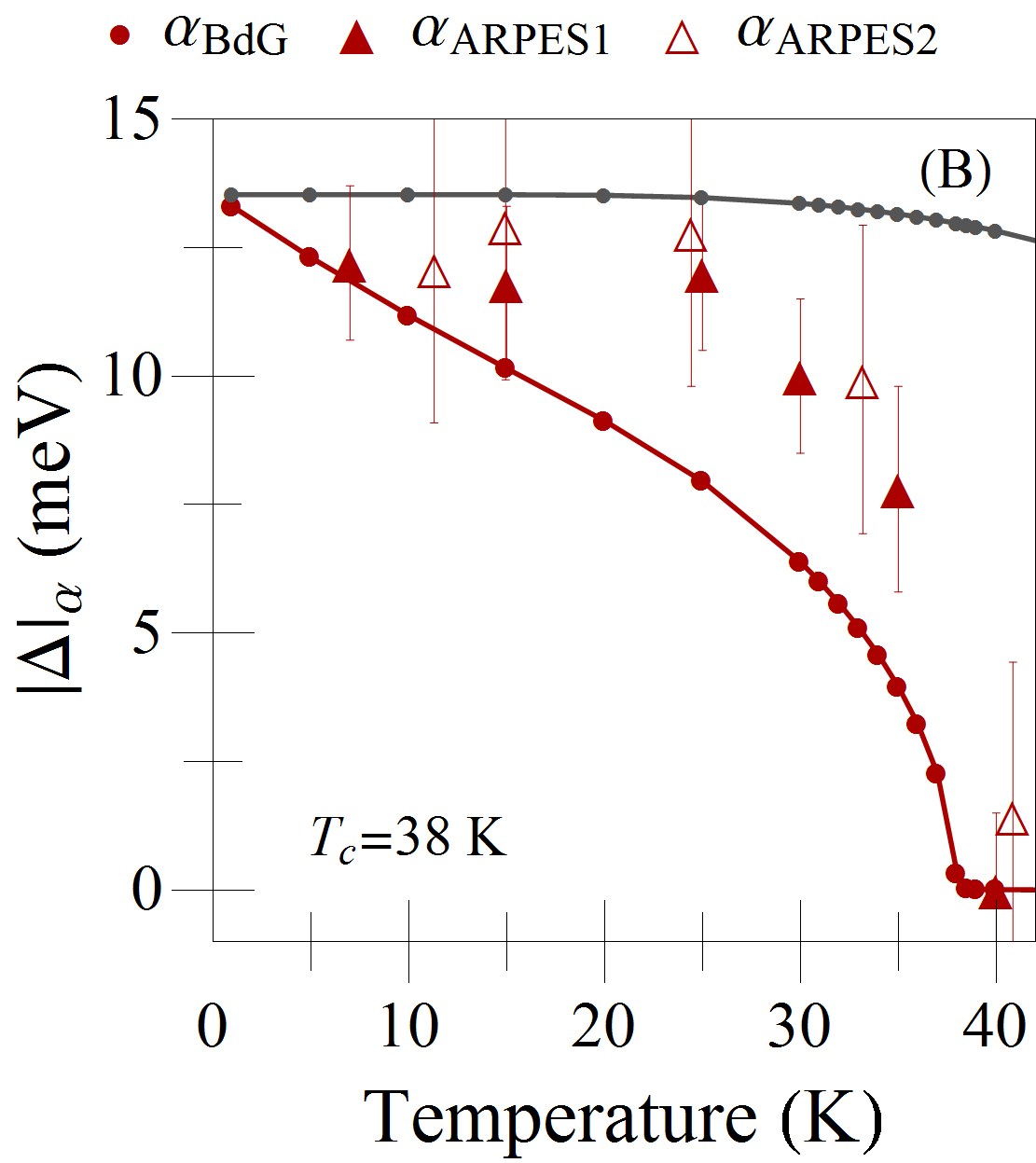}%
\end{subfigure}
\begin{subfigure}[b]{0.24\textwidth}
\includegraphics[width=\textwidth]{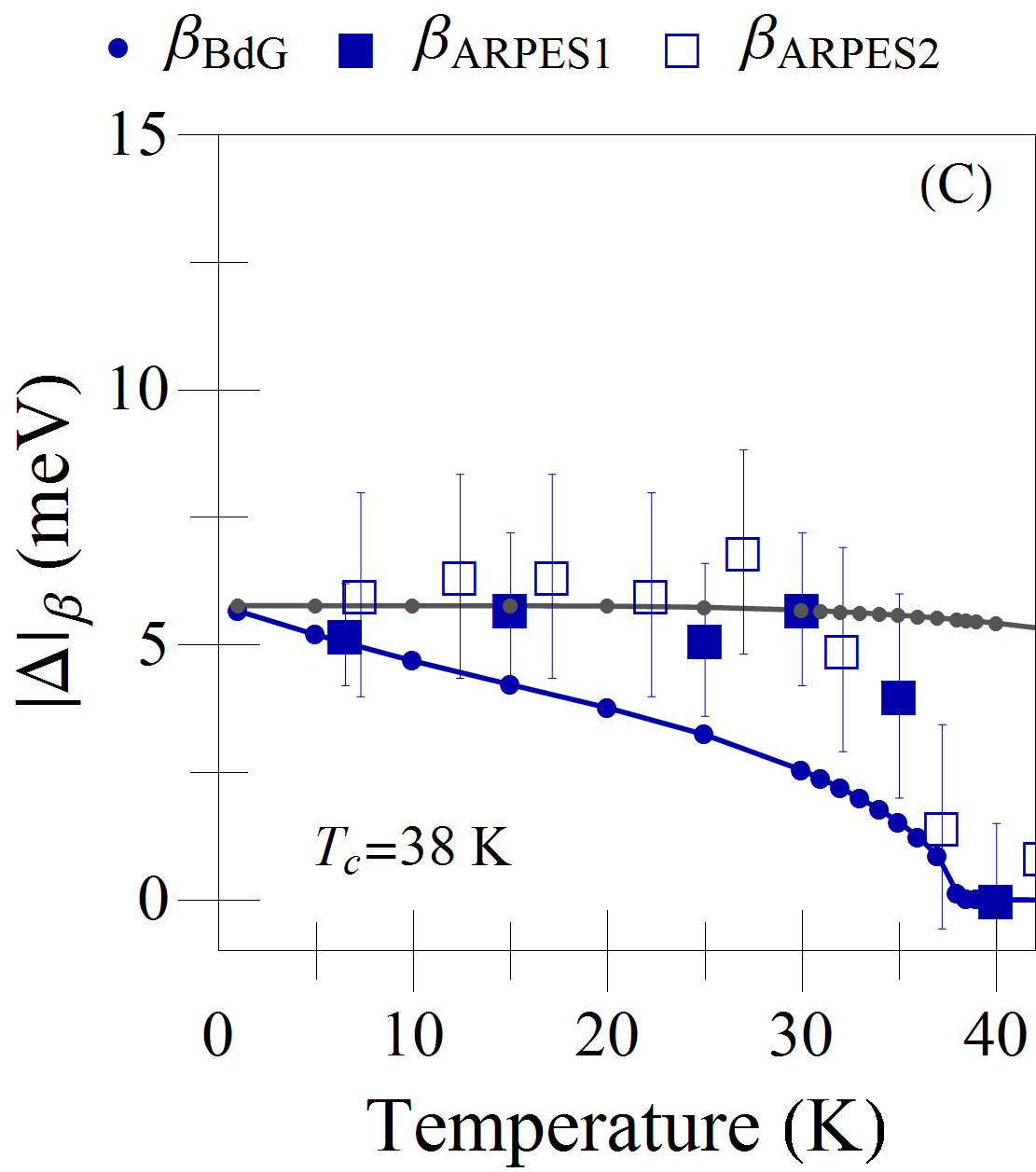}%
\end{subfigure}
\begin{subfigure}[b]{0.24\textwidth}
\includegraphics[width=\textwidth]{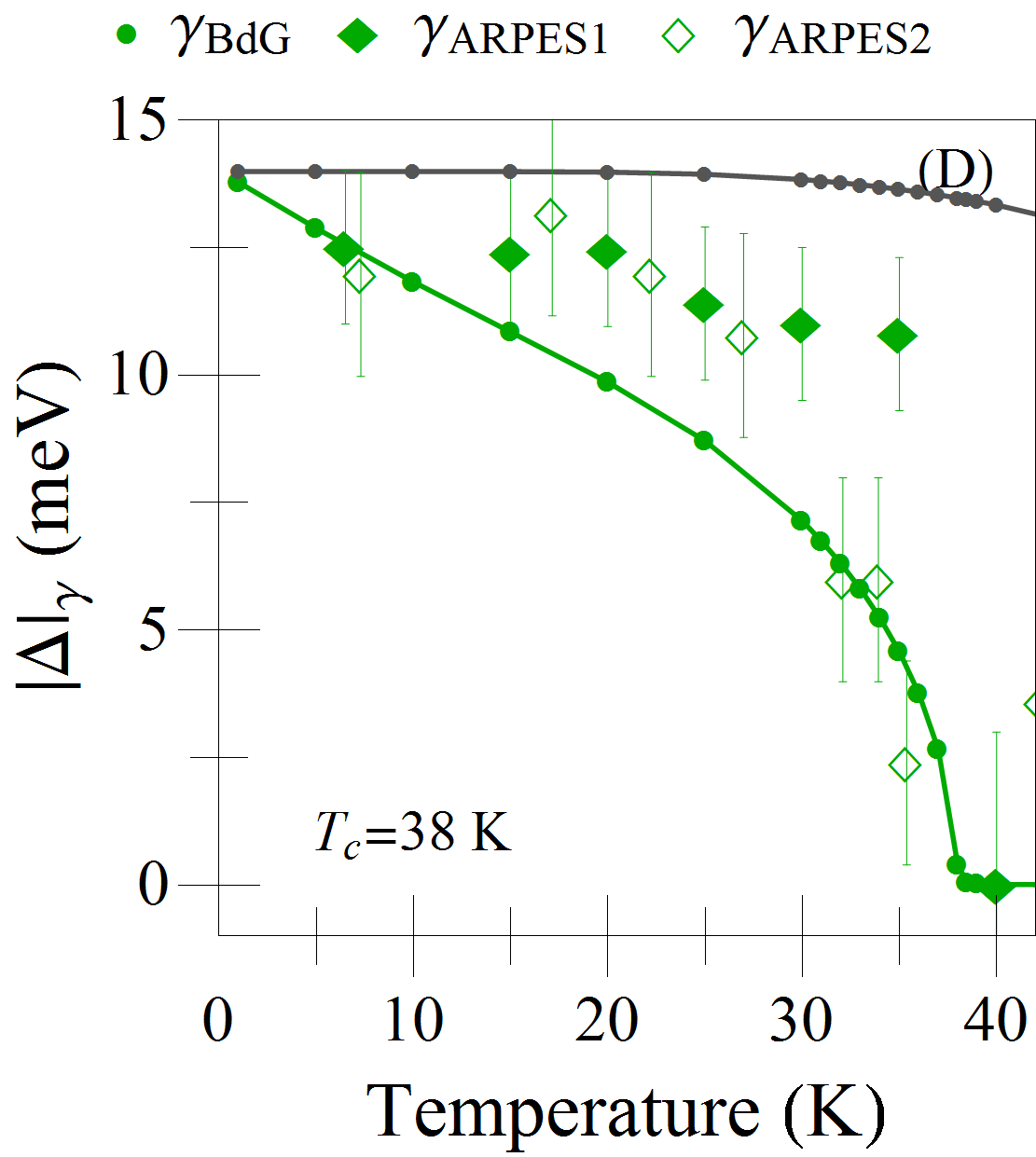}%
\end{subfigure}
\caption{
Temperature dependence of (a) the jointly obtained SC gaps $\Delta_\nu(T)$ from multi-band BdG theory (solid lines) in comparison to references \cite{Ding2008} (empty points) and \cite{Wray2008} (solid points) for Ba$_{0.6}$K$_{0.4}$Fe$_2$As$_2$ with $T_c=38$ K and $N=8$. We show the three sets of data corresponding to (b) $\nu=\alpha$, (c) $\nu=\beta$, and (d) $\nu=\gamma$  separately for clarity. The grey curves correspond to the case of a constant potential, which is included for comparison. \label{tc38}
}%
\end{figure*}

Now we apply multi-band BdG theory to model three compounds of 
Ba$_{1-x}$K$_{x}$Fe$_2$As$_2$ with $T_c=23$, $32$ and $38$ K
to reproduce the SC gaps measured by different techniques.
PCAR techniques \cite{Samuely2009a,Szabo2009} find two SC bands, whereas ARPES \cite{Ding2008} detects SC gaps in three bands 
or pockets in the Brillouin zone. It is
possible that the bigger gap observed by PCAR may correspond to the $\alpha$ 
and $\gamma$ bands from ARPES, since they have about the same intensity. 

For each compound we solve
a three-band $6N^2\times 6N^2$  BdG matrix in a square lattice with periodic boundary
conditions.
Following the LDA calculations \cite{Singh2008,Ma2008,Xu2008a} 
and the derived bands by ARPES\cite{Ding2011} we can obtain an estimation of the 
inter-band hoppings shown in the first three columns of table \ref{tab:tablehop}.
For instance, there is very little overlap between and no
intersection between $\alpha$ and $\gamma$ and so we take $t_{\alpha\gamma}=0$ and larger inter-band hoppings to reflect band intersections between 
the $\alpha$ and $\beta$ and $\beta$ and $\gamma$ pockets \cite{Sato2009,Ding2011}, which are shown in the last three columns of table \ref{tab:tablehop}. 

To model the experiment results we perform 
calculations with these hoppings listed in table \ref{tab:tablehop}, used for all three compounds.
The charge distribution in each pocket $(\rho_\alpha,\rho_\beta,\rho_\gamma)=(0.192,0.084,0.12)$ are proportional to specific Fermi areas \cite{Sato2009}, and Fermi velocities \cite{Ding2008,Evtushinsky2009a}. Even if the 
system has a disordered electronic density, the  average gap is the same of that of a homogeneous one with the same average density. Furthermore, ARPES and PCAR 
measure only the average gap.

\begin{table}
\centering
\captionsetup{justification=raggedright,
singlelinecheck=false
}
\caption{\label{tab:tablehop}First and second nearest neighbour intra $t_{\nu}$ and inter-band hoppings $t_{\mu\nu}$ in meV used in the present calculations. Intra-band hoppings were extracted from ARPES and inter-band hoppings closely agree with band dispersion \cite{Ding2011}. The parameters of this table are used for all three compound discussed in this paper.}
\begin{ruledtabular}
\begin{tabular}{
lllllll
}
Hoppings (meV) & $t_{\alpha}$ & $t_{\beta}$ & $t_{\gamma}$ & $t_{\alpha\beta}$ & $t_{\alpha\gamma}$ & $t_{\beta\gamma}$  \\
\hline
$1^\mathrm{st}$ & $160$ & $13$ & $380$ & $165$ & $0$ & $140$\\
$2^\mathrm{ond}$ & $-52$ & $42$ & $80$ & $0$ & $0$ & $0$\\
\end{tabular}
\end{ruledtabular}
\end{table}

\begin{table}
\centering
\captionsetup{justification=raggedright,
singlelinecheck=false
}
\caption{\label{tab:tablephen}Phenomenological values of the electronic phase transition temperature $T_\mathrm{PS}$ and potential $|V_0|$ in equation \eqref{varying_temp} used to model Ba$_{1-x}$K$_{x}$Fe$_2$As$_2$ with different $T_c$'s.}
\begin{ruledtabular}
\begin{tabular}{
llll
}
Compound & $T_c$ (K) & $T_\mathrm{PS}$ (K) & $|V_0|$ (meV) \\
\hline
Ba$_{0.6}$K$_{0.4}$Fe$_2$As$_2$ & $38$  & $165$ &$-400$\\
Ba$_{1-x}$K$_{x}$Fe$_2$As$_2$ & $32$  & $130$ &$-360$\\
Ba$_{1-x}$K$_{x}$Fe$_2$As$_2$ & $23$  & $80$ &$-300$\\
\end{tabular}
\end{ruledtabular}
\end{table}

To obtain the SC gaps of different Ba$_{1-x}$K$_{x}$Fe$_2$As$_2$ compounds
with distinct $T_c$'s we 
used the band parameters discussed before and 
estimate the two-body attractive potential  in equation \eqref{varying_temp} as listed in table \ref{tab:tablephen}. The $\Delta_\nu(T)$ results for the compound with $T_c=23$ K is depicted on the left panel of figure \ref{tc2332}. 
The coloured solid lines are the theoretical BdG curves obtained from \eqref{swavegap}. We observe that PCAR \cite{Szabo2009,Samuely2009a} (black triangles) scales 
more closely with the BdG $\alpha$/$\gamma$ bands.
In the case of the compound with $T_c=32$ K we show also experimental points from ARPES \cite{Evtushinsky2009} in comparison to the theoretical BdG curves on the right panel of figure \ref{tc2332}. Experimental points for the $\beta$ bands were unavailable and estimated under 4 meV \cite{Evtushinsky2009}, consistent with the blue $\beta$ BdG curve.

The results for Ba$_{0.6}$K$_{0.4}$Fe$_2$As$_2$ with $T_c=38$ K are shown in figure \ref{tc38},  with all bands (left panel) and with each band shown separately for clarity. Two different sets of ARPES points are displayed denoted by ARPES1 referring to \cite{Ding2008} and ARPES2 according to \cite{Wray2008}. It is interesting to note that the measured  band's correspondent SC gaps are 
reproduced by a temperature-dependent attractive potential within a three-band BdG theory. This is indicated by the grey curves in figure \ref{tc38}, which correspond to the case where a constant potential that gives the same $\Delta_{\nu}(T=0)$ was used.

Despite the real and $k$-space complexity of the system, 
we obtain a reasonable agreement with the experimental values, using a single attractive potential (equation \eqref{varying_temp}) for the three bands for each compound.
The misfit of the experimental points with respect to the theoretical BdG curves may be due to differences in the tight binding hoppings and, more importantly,
in the finite size used
in the calculations with respect to the real system. Our matrix is only
$384 \times 384$ and it is solved by exact diagonalization.
New calculations using a more powerful method \cite{Covaci2010,Weisse2005b} that allows investigation of much larger arrays are being presently considered. However, we believe that the essential properties of these compounds are captured by the present calculations.

\section{Conclusion}

We have discussed along the paper the evidence of electronic disorder 
complexity in real and $k$-space of the
Ba$_{1-x}$K$_{x}$Fe$_2$As$_2$ compounds.  A possible electronic 
phase separation or a nematic order together
with the multi-orbital band structure make 
this system quite unusual. As expected the description to its SC multi-gap properties
is not a simple task, unlike MgB$_2$ with two bands and two conventional
BCS-like gaps \cite{Wang2001}.

The SC multi-gap structure revealed by ARPES \cite{Ding2008,Ding2011} 
with three different gaps on distinct locations
of the Brillouin zone, but with a single $T_c$, in light of the  
two band theory of Suhl et al \cite{Suhl1959}, is an indication of inter-band
interaction or hybridization. This was confirmed by 
self-consistent calculations with the BdG theory with inter-band hoppings. These hoppings describe phenomenologically the interaction between bands, mediated by the hybridized iron orbitals.

The anisotropy in real space was also taken into consideration.
In analogy with the cuprate superconductors which displayed many
indications of charge disorder,\cite{Comin2014,DaSilvaNeto2014}
we also use the observed charge disorder or nematic order as the origin
of a two-body attractive pair potential with a temperature dependence derived
from a GL theory.

We conclude that the multi-band scattering via multi-orbital 
hybridizations and in plane charge disorder are well known important
properties of high complexity of these materials that are essential
to quantify the pairing amplitudes and the overall SC properties.

\section*{Acknowledgments}
We gratefully acknowledge partial financial aid from Brazilian agencies CAPES, FAPERJ and CNPq.










\bibliography{jop}

\end{document}